\documentclass[%
 reprint,
 amsmath,amssymb,
 aps,
]{revtex4-2}

\usepackage{graphicx}
\usepackage{dcolumn}
\usepackage{bm}
\usepackage{bbm}

\usepackage[utf8]{inputenc}  
\usepackage[T1]{fontenc}     

\usepackage{amsmath}
\usepackage{amssymb}
\usepackage{graphicx}
\usepackage{mathrsfs}
\usepackage{amsfonts}
\usepackage{amsthm}
\usepackage{color}
\usepackage{txfonts}
\usepackage[colorlinks=true,citecolor=blue,linkcolor=blue,urlcolor=blue,anchorcolor=blue]{hyperref}%
\usepackage{pifont}
\hypersetup{colorlinks=true,citecolor=blue,linkcolor=blue,urlcolor=blue}
\usepackage{lineno} 	
\providecommand{\U}[1]{\protect\rule{.1in}{.1in}}
\makeatletter
\@ifundefined{textcolor}{}
{
\definecolor{BLACK}{gray}{0}
\definecolor{WHITE}{gray}{1}
\definecolor{RED}{rgb}{1,0,0}
\definecolor{GREEN}{rgb}{0,1,0}
\definecolor{BLUE}{rgb}{0,0,1}
\definecolor{CYAN}{cmyk}{1,0,0,0}
\definecolor{MAGENTA}{cmyk}{0,1,0,0}
\definecolor{YELLOW}{cmyk}{0,0,1,0}
}

\begin{document}


\title{Emergent Skyrmion Hall Effect in $d$-wave Altermagnets at Finite Temperature}

\author{Tingting Liu$^{1}$}
\email[Corresponding author: ]{tingtingliu@lcu.edu.cn}
\author{Bingyu Sun$^{1}$}
\author{Fengyue Zhu$^{1}$}
\author{Zhihui Zhang$^{1}$}
\author{Peiyu Zhang$^{1}$}

\author{Yang Liu$^{2}$}
\email[Corresponding author: ]{dzly1019@163.com}

\author{Minghui Qin$^{3}$}
\email[Corresponding author: ]{qinmh@scnu.edu.cn}

\affiliation{$^{1}$School of Physics Science and Information Technology, Liaocheng University, Liaocheng 252000, China}
\affiliation{$^{2}$School of Physics and State Key Laboratory of Electronic Thin Films and Integrated Devices, University of Electronic Science and Technology of China, Chengdu 610054, China}
\affiliation{$^{3}$Guangdong Provincial Key Laboratory of Quantum Engineering and Quantum Materials and Institute for Advanced Materials, South China Academy of Advanced Optoelectronics, South China Normal University, Guangzhou 510006, China}

\begin{abstract}
 Altermagnets combine compensated magnetic order with unconventional symmetry-dependent responses, offering a promising platform for spintronic applications. Here, we show that a voltage-controlled magnetic-anisotropy gradient drives altermagnetic (ATM) skyrmions in a rectilinear, Hall-free motion in the absence of thermal fluctuations, owing to their strongly compensated gyrotropic response. Thermal magnons qualitatively modify this behavior by increasing the longitudinal drag through magnon--skyrmion scattering and generating a transverse reaction force through handedness-dependent skew scattering. Owing to the anisotropic altermagnetic magnon band structure, the relative transport weights of the two magnon handednesses are interchanged between propagation along the $x$ and $y$ directions, resulting in opposite signs of the Hall angle for driving along the $x$ and $y$ directions. By contrast, along the high-symmetry direction, the two magnon handednesses remain degenerate and their transverse contributions cancel, preserving Hall-free motion even at finite temperature. We thus uncover a thermally emergent anisotropic skyrmion Hall effect whose direction-dependent magnitude and sign originate from the intrinsic symmetry-dependent magnon spectrum, making it a generic finite-temperature dynamical feature of ATM skyrmions. Our results establish a low-power route toward electrically controlled and thermally tunable ATM skyrmion transport.
\end{abstract}

\maketitle

\section{\label{sec:level1}INTRODUCTION}
 Altermagnets, a recently identified class of magnetic materials distinct from conventional ferromagnets and antiferromagnets \cite{PhysRevX.12.031042,Jiang2025}, have attracted significant attention and are considered promising candidates for advanced spintronic applications \cite{Song2025,y9q4-13fw, PhysRevLett.129.137201, PhysRevLett.130.036702}. A defining feature of altermagnets is the emergence of nonrelativistic spin splitting in the electronic band structure despite their compensated collinear magnetic order. This property enables them to generate time-reversal-symmetry-breaking magnetotransport phenomena similar to those observed in ferromagnets \cite{Gomonay2024}. At the same time, altermagnets retain several advantages commonly associated with antiferromagnets, including ultrafast terahertz spin dynamics, robustness against external perturbations, and nearly vanishing stray magnetic fields \cite{PhysRevLett.131.256703}. These remarkable characteristics originate from the special interplay between crystal symmetry and spin symmetry, which gives rise to an alternating non-relativistic spin polarization \cite{PhysRevLett.134.176401,PhysRevLett.133.196701}. The unconventional spin symmetries of altermagnets can imprint distinct real-space structures and dynamics on their spin textures. In particular, sublattice canting generates finite local magnetic moments \cite{Gomonay2024}, giving rise to stripe domains and skyrmions with emergent magnetic multipoles \cite{PhysRevLett.133.196701,PhysRevLett.134.176401}. Such canted spin profiles exhibit unconventional dynamical responses, including anisotropic Walker breakdown \cite{Gomonay2024} and a skyrmion Hall effect \cite{PhysRevLett.133.196701,PhysRevLett.134.176401}, which distinguish them from their counterparts in conventional antiferromagnets.
 
 Magnetic skyrmions are particle-like topological spin textures commonly stabilized in chiral magnets by broken inversion symmetry and the Dzyaloshinskii-Moriya interaction (DMI) \cite{PhysRevLett.120.117201, Jin_2022, https://doi.org/10.1002/adma.202513067, https://doi.org/10.1002/adma.202270090, Caretta2018,Wang2020}. Their nanoscale dimensions and topological robustness make them promising information carriers, motivating extensive studies of their creation, annihilation, and manipulation \cite{PhysRevB.102.054419, WANG2023113484, Nagaosa2013, doi:10.1126/science.aaa1442, Peng2017, 10.1063/1.4967006, PhysRevApplied.12.044031, PhysRevB.104.174421,PhysRevLett.122.057204,PhysRevB.92.020403,PhysRevLett.111.067203,PhysRevApplied.18.024062}. For many candidate altermagnets, however, current-driven schemes such as spin-transfer torque can be ineffective or inefficient because of their insulating or semiconducting character, while Joule heating further limits their energy efficiency \cite{10.1063/5.0056259,10.1063/5.0250430}. Voltage-controlled magnetic anisotropy (VCMA) gradients offer a current-free strategy for driving altermagnetic (ATM) skyrmions by converting an applied electric field into a spatially varying magnetic energy landscape. Electric-field control of skyrmions and domain walls through magnetic-anisotropy modulation has been demonstrated theoretically and experimentally in a range of magnetic heterostructures \cite{PhysRevB.101.024414,PhysRevB.98.134448,PhysRevB.98.024421,chiba2011electrical,PhysRevB.96.220412}, suggesting VCMA as a promising approach for manipulating ATM skyrmions. Moreover, the emergent magnetic quadrupole of ATM skyrmions can modify their collective dynamics relative to those in conventional antiferromagnetic order \cite{y9q4-13fw,PhysRevLett.134.176401, PhysRevLett.133.196701}, warranting a dedicated investigation under VCMA driving. 
 
 \setlength{\parskip}{0pt} 
 Beyond the modification of collective dynamics by the emergent magnetic quadrupole, thermal excitations introduce an additional degree of complexity into ATM skyrmion motion. In conventional ferromagnetic, antiferromagnetic, and ferrimagnetic systems, thermal fluctuations and thermally driven magnons are known to induce stochastic motion, additional dissipation, and magnon-mediated forces on magnetic textures \cite{PhysRevLett.111.067203, PhysRevLett.116.147203, PhysRevResearch.2.013293}. Altermagnets provide a distinct setting because their compensated magnetic order coexists with strongly anisotropic and handedness-dependent magnon spectra \cite{PhysRevLett.131.256703, PhysRevB.111.L020407, PhysRevB.108.L180401}. Thermal magnons may therefore couple to ATM skyrmions in a direction-dependent manner and generate dynamical responses beyond those expected in conventional compensated magnets. However, how thermal fluctuations and anisotropic magnon transport jointly influence driven ATM skyrmion dynamics remains largely unexplored. 
 
 \setlength{\parskip}{0pt} 
 In this work, we investigate the dynamics of ATM skyrmions driven by a VCMA gradient using collective-coordinate theory and atomistic spin simulations. In the absence of thermal fluctuations, the strongly compensated gyrotropic response leads to rectilinear, Hall-free motion, with the longitudinal velocity controlled by the anisotropy gradient, Gilbert damping, and background anisotropy. At finite temperatures, thermally excited magnons introduce additional longitudinal drag and generate a transverse force through handedness-dependent skew scattering. Owing to the anisotropic altermagnetic magnon spectrum, the relative transport weights of the two magnon handednesses are interchanged between propagation along the $x$ and $y$ directions, producing skyrmion Hall angles of opposite sign. Along the high-symmetry direction, however, the two magnon branches remain degenerate and their transverse contributions cancel, preserving Hall-free motion. These results reveal a thermally emergent anisotropic skyrmion Hall effect that constitutes a generic finite-temperature dynamical feature of ATM skyrmions and may serve as a dynamical signature of altermagnetic order.

    \begin{figure}
    	\centering
    	\includegraphics[width=.9\columnwidth]{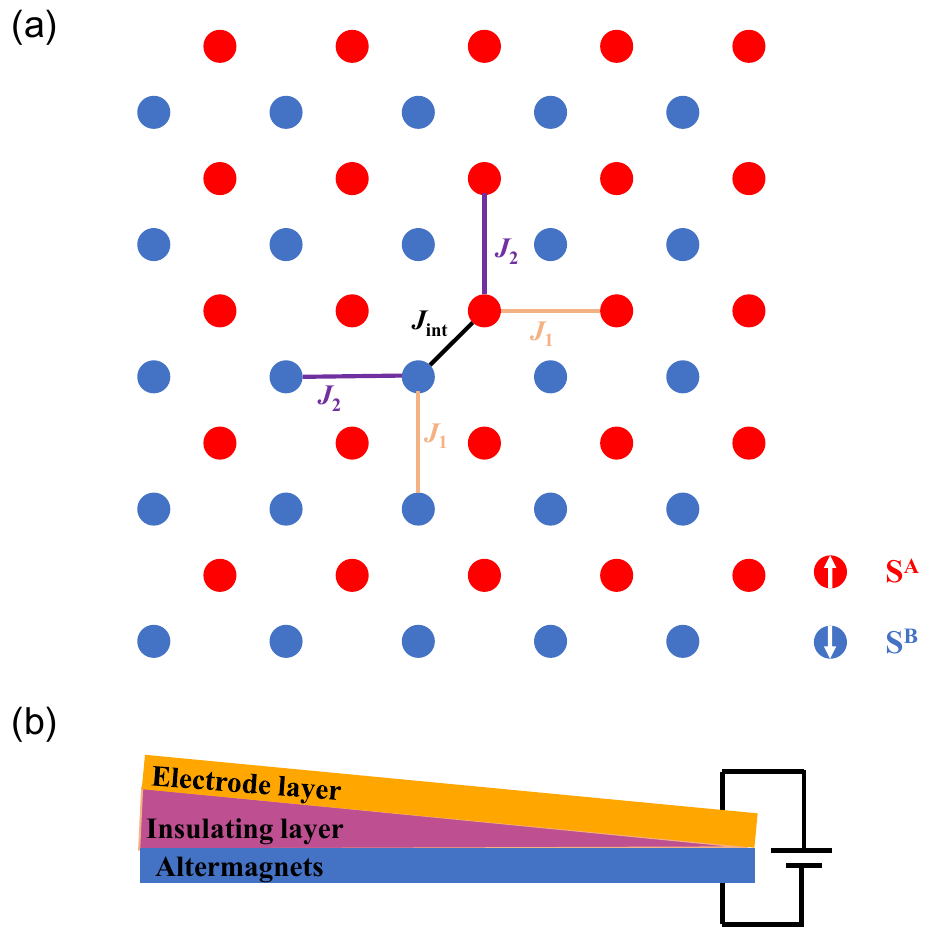}
    	\caption{(a) Two sublattices of magnetic atoms in different layers represented by red and blue circles, respectively. $\mathbf{S}^n  (n = A, B)$ denotes the spin direction. To clearly illustrate the distinct exchange interactions between the two sublattices, one atomic layer has been shifted by $\sqrt{2}a/2$ along the diagonal direction. (b) The structure of the voltage-controlled magnetic anisotropy
    device.}
    	\label{fig1}
    \end{figure}

\section{\label{sec:level2}MODEL AND METHODS}

 We consider a two-sublattice altermagnetic model with anisotropic intrasublattice exchange interactions, as shown in Fig.~\ref{fig1}. The corresponding $d$-wave altermagnetic Hamiltonian is given by
    \begin{subequations}\label{eq1}
    \begin{align*}
    { H}=&-\sum_{i,j} [ J_1 \mathbf{S}_{i,j}^A \cdot \mathbf{S}_{i+1,j}^A + J_2 \mathbf{S}_{i,j}^B \cdot \mathbf{S}_{i+1,j}^B \\
    &+ J_2 \mathbf{S}_{i,j}^A \cdot \mathbf{S}_{i,j+1}^A + J_1 \mathbf{S}_{i,j}^B \cdot \mathbf{S}_{i,j+1}^B
    + J_\text{int} \mathbf{S}_{i,j}^A \cdot \mathbf{S}_{i,j}^B\\
    &- D_0 (\mathbf{S}_{i,j}^A \times \mathbf{S}_{i,j+1}^A + \mathbf{S}_{i,j}^B \times \mathbf{S}_{i,j+1}^B) \cdot \mathbf{x} \\
    &+ D_0 (\mathbf{S}_{i,j}^A \times \mathbf{S}_{i+1,j}^A + \mathbf{S}_{i,j}^B \times \mathbf{S}_{i+1,j}^B) \cdot \mathbf{y} \\
    &+ K_z (\mathbf{S}_{i,j}^A \cdot \mathbf{z})^2 + K_z (\mathbf{S}_{i,j}^B \cdot \mathbf{z})^2].  \tag{1}
    \end{align*}
    \end{subequations}
 Here, $J_{1,2}>0$ represents the intralayer anisotropic exchange coupling, $J_\text{int}<0$ is the interlayer antiferromagnetic coupling, and $D_0$ and $K_z$ denote the interfacial DMI and magnetic anisotropy coefficients, respectively. $\mathbf{S}_{i,j}^A$ and $\mathbf{S}_{i,j}^B$ are the normalized spins on sites $(i,j)$ of sublattices $A$ and $B$, respectively. For the altermagnetic bilayer model considered here, we assume two crystal mirror planes containing each interlayer bond, which forbid DMI on these bonds according to Moriya’s symmetry rules \cite{PhysRev.120.91}. By introducing the net spin $\mathbf{m}=(\mathbf{S}^A+\mathbf{S}^B)/2$ and N\'eel vector $\mathbf{n}=(\mathbf{S}^A-\mathbf{S}^B)/2$, we rewrite the model in the continuum limit for slowly varying spin textures \cite{PhysRevB.93.104408, PhysRevLett.133.196701}. Retaining the leading contributions in spatial gradients and the small magnetization $\mathbf{m}$, while neglecting higher-order terms, we obtain the continuum Hamiltonian density
    \begin{subequations}\label{eq2}
    \begin{align*}
    {\cal H}=&\frac{A_0}{2} \mathbf{m}^2 + A_1 (\nabla \mathbf{n})^2 + A_2 (\partial_x \mathbf{m} \cdot \partial_x \mathbf{n} - \partial_y \mathbf{m} \cdot \partial_y \mathbf{n}) \\
    &+ D n_z \nabla \cdot \mathbf{n} - D (\mathbf{n} \cdot \nabla) n_z - K n_z^2, \tag{2}
    \end{align*}
    \end{subequations}
 where ${A_0 = -{4J_\text{int}}/{a^3}}$, ${A_1 = {J_1 + J_2}/{a}}$, ${A_2 = {J_2 - J_1}/{a}}$, ${D = {2D_0}/{a^2}}$ and ${K = {2K_z}/{a^3}}$ are the homogeneous, inhomogeneous exchange, anisotropic exchange, DMI and magnetic anisotropy energies, respectively. Here, $a$ represents the lattice constant. Different from the antiferromagnets, the altermagnetic exchange term $ A_2 (\partial_x \mathbf{m} \cdot \partial_x \mathbf{n} - \partial_y \mathbf{m} \cdot \partial_y \mathbf{n}) $ occurs in the above Hamiltonian, it breaks the symmetry of conventional antiferromagnets. In this work, the magnetic anisotropy is assumed to vary linearly along a chosen inplane direction, $K(\xi)=K_0-\xi \Delta K_\xi$, where $\xi$ is the spatial coordinate along the gradient direction, $K_0$ denotes the background uniaxial anisotropy, and $\Delta K_\xi$ is the corresponding anisotropy-gradient magnitude. Starting from the coupled Landau--Lifshitz--Gilbert equations for the two sublattices \cite{PhysRevLett.106.107206}, the equations of motion for the magnetization $\mathbf{m}$ and N\'eel vector $\mathbf{n}$ read
    \begin{subequations}\label{eq3}
    \begin{align*}
    {\dot{\mathbf{m}}}=-\frac{1}{2s}(\mathbf{m}\times\mathbf{f_m}+\mathbf{n}\times\mathbf{f_n}) + \alpha(\mathbf{m}\times\dot{\mathbf{m}}+\mathbf{n}\times\dot{\mathbf{n}}), \tag{3a}\\
    {\dot{\mathbf{n}}}=-\frac{1}{2s}
    (\mathbf{m}\times\mathbf{f_n}+\mathbf{n}\times\mathbf{f_m}) + \alpha(\mathbf{m}\times\dot{\mathbf{n}}+\mathbf{n}\times\dot{\mathbf{m}}). \tag{3b}
    \end{align*}
    \end{subequations}
 where $\alpha$ is the Gilbert damping constant, $s = \mu_s/\gamma$ is the spin angular momentum with the atomic magnetic moment $\mu_s$ and gyromagnetic ratio ${\gamma}$. $\mathbf{f_n} = -{\delta\cal H}/\delta\mathbf{n}$ and $\mathbf{f_m} = -{\delta\cal H}/\delta\mathbf{m}$ denote the effective fields of $\mathbf{n}$ and $\mathbf{m}$, respectively. Within the exchange approximation, we combine Eqs.~(\ref{eq3}a) and (\ref{eq3}b) and substitute the resulting expression for $\mathbf{m}$ back into the equations of motion \cite{PhysRevLett.106.107206}, obtaining
    \begin{subequations}\label{eq4}
    \begin{equation}
    {\rho\ddot{\mathbf{n}}}-2s\mathbf{n}\times\dot{\mathbf{A}}=(\mathbf{n}\times\mathbf{f_n})\times\mathbf{n}-2s\alpha\dot{\mathbf{n}},\tag{4}
    \end{equation}
    \end{subequations}
 where $\rho=-s^2/J_\text{int}$ is the inertia of the N\'eel vector, and $\mathbf{A} = (J_{1}-J_{2})a^2/J_\text{int}\,\mathbf{n} \times \allowbreak \bigl[(\partial_x^2\mathbf{n}-\partial_y^2\mathbf{n}) \times \mathbf{n}\bigr]$ is the magnetic quadrupole induced by $C_{4}\mathcal{T}$ symmetry in $d$-wave altermagnets. Furthermore, we assume that the moving skyrmion retains a fixed shape and can be described by its guiding center $\mathbf{R}(t)$ as $\mathbf{n}=\mathbf{n}_0[\mathbf{r}-\mathbf{R}(t)]$ \cite{PhysRevLett.30.230}, where $\mathbf{n}_0$ is the equilibrium skyrmion profile in the absence of a magnetic anisotropy gradient. We then introduce a weak magnetic anisotropy gradient, assumed not to appreciably deform the skyrmion profile. Substituting the rigid-translation ansatz into Eq.~(\ref{eq4}) gives $\dot{\mathbf n}=-\dot R_j\partial_j\mathbf n$ and $\ddot{\mathbf n}\approx-\ddot R_j\partial_j\mathbf n$, where terms quadratic in the skyrmion velocity are neglected. Projecting Eq.~(\ref{eq4}) onto the translational modes and integrating over space, with the boundary terms vanishing for a localized texture, yields the collective equation of motion
    \begin{subequations}\label{eq5}
    \begin{equation}
    {\overleftrightarrow{\mathcal{M}} \ddot{\mathbf{R}} + 2s\overleftrightarrow{\mathcal{G}} \dot{\mathbf{R}}+2\mathcal{K}-2s\alpha\overleftrightarrow{\mathcal{D}}\dot{\mathbf{R}}=0 }.\tag{5}
    \end{equation}
    \end{subequations}
 Here, $\mathcal{M}_{ij} = \rho \int \bigl( \partial_i \mathbf{n} \cdot \allowbreak \partial_j \mathbf{n} \bigr) d \mathbf{r}$ is the effective ATM skyrmion mass of a tensor form $(i,j=x,y)$ and $\mathcal{D}_{ij} = \int \left( \partial_i \mathbf{n} \cdot \partial_j \mathbf{n} \right) d\mathbf{r}$ is the dissipative tensor. $\mathcal{K} = (\Delta K_x\int \left( 1 - n_z^2 \right) d\mathbf{r},~\Delta K_y\int \left( 1 - n_z^2 \right) d\mathbf{r})$ is a VCMA-induced force vector. $\ddot{\mathbf{R}}(t)$ and $\dot{\mathbf{R}}(t)=(v_x, v_y)$ are the ATM skyrmion acceleration and velocity, respectively. $\mathcal{G}_{ij} = \int ( \mathbf{n} \times \partial_i\mathbf{A} \cdot \partial_j \mathbf{n} ) d\mathbf{r}$ is a tensor associated with the magnetic quadrupole $\mathbf{A}$. In addition, we find that the intrinsic $C_{4}\mathcal{T}$ symmetry requires $\mathcal{G}_{ij}=0$. Solving Eq.~(\ref{eq5}) in the steady-state limit yields the ATM skyrmion velocity components
    \begin{subequations}\label{eq6}
    \begin{equation}
            v_x=\frac{\mathcal{K}_x}{s\alpha \mathcal{D}}, \tag{6a}
    \end{equation}
    \vspace{-1.5em} 
    \begin{equation}
            v_y=\frac{\mathcal{K}_y}{s\alpha \mathcal{D}}, \tag{6b}
    \end{equation}
    \end{subequations}
 with $\mathcal{D}=\mathcal{D}_{xx}=\mathcal{D}_{yy}$. Equation (\ref{eq6}) shows that the ATM skyrmion exhibits linear motion, moving toward regions of reduced magnetic anisotropy without Hall angle at zero temperature. For $\mathcal{G}_{ij}=0$, Eq.~(\ref{eq6}) recovers the conventional AFM-skyrmion result under a magnetic anisotropy gradient \cite{PhysRevB.98.134448}, with Hall-free motion arising from the cancellation of the transverse gyrotropic forces. Conversely, applying an opposite voltage results in an inverted $\Delta{K}_\xi$, thereby propelling the ATM skyrmion in the opposite direction. Moreover, the velocity is proportional to $\Delta{K}_\xi$ and $1/\alpha$.

 At finite temperatures, thermally excited magnons form a bath that exchanges momentum and energy with the skyrmion through magnon--skyrmion scattering \cite{PhysRevE.48.4037, PhysRevB.89.064412,PhysRevB.90.094423}. We account for the resulting dissipative backaction by replacing $s\alpha\mathcal D$ with $\Gamma_{\mathrm{eff}}(T)=s\alpha\mathcal D+\eta T$, where $\eta$ is a phenomenological friction coefficient characterizing the temperature-induced dissipative contribution \cite{PhysRevLett.127.047203}. In addition to this longitudinal drag, asymmetric magnon skew scattering can transfer transverse momentum to the skyrmion. Although the reaction forces cancel for a stationary skyrmion in an isotropic thermal bath, skyrmion motion renders the relative magnon flux anisotropic in its rest frame, thereby producing a velocity-dependent transverse force.

 To describe the backaction of thermal magnons, we resolve the magnon bath into two branches of opposite handedness, labeled by $\nu=\pm$. Their energies and group velocities are
    \begin{equation}
        \varepsilon_\nu(\mathbf k)=\hbar\omega_\nu(\mathbf k),
        \qquad
        \mathbf v_{g,\nu}(\mathbf k)
        =
        \frac{1}{\hbar}
        \nabla_{\mathbf k}\varepsilon_\nu(\mathbf k).
        \label{eq7}
    \end{equation}
 Writing $\mathbf k=k(\cos\varphi,\sin\varphi)$, where $k=|\mathbf k|$ and $\varphi$ is the polar angle measured from the $k_x$ axis, the interchange of the two branches under a rotation by $\pi/2$ is expressed as $\varepsilon_\nu(k,\varphi+\pi/2) = \varepsilon_{-\nu}(k,\varphi)$. The angular dependence must therefore be retained when combining the contributions from the two branches, reflecting the $C_4\mathcal{T}$ symmetry of the altermagnet and the resulting spin splitting and interchange of the chiral magnon branches under a $90^\circ$ rotation. With the stochastic field included \cite{PhysRevLett.117.217201, PhysRevLett.127.047203}, the equilibrium magnon occupation follows the Rayleigh–Jeans distribution,
    \begin{equation}
        n_\nu^{\mathrm{cl}}(\mathbf k,T)
        =
        \frac{k_{\mathrm B}T}{\varepsilon_\nu(\mathbf k)}.
    \end{equation}
 Consequently, \(n_\nu^{\mathrm{cl}}(k,\varphi+\pi/2)=n_{-\nu}^{\mathrm{cl}}(k,\varphi)\), indicating an interchange of the branch occupations between propagation directions.

  In compensated magnetic textures, the two oppositely polarized magnon branches can undergo transverse deflections in opposite directions \cite{PhysRevB.99.224433, PhysRevB.104.054419}. Their contributions to the skyrmion transverse force are therefore determined by the signed skew-scattering response. For a skyrmion moving with velocity $\mathbf v=\dot{\mathbf R}$, the comoving-frame energy is $\varepsilon_\nu' =\varepsilon_\nu-\hbar\mathbf k\cdot\mathbf v$. Expanding the laboratory-frame equilibrium occupation $n^{\mathrm{cl}}(\varepsilon_\nu' +\hbar\mathbf k\cdot\mathbf v,T)$ about $n^{\mathrm{cl}}(\varepsilon_\nu',T)$ gives
    \begin{equation}
        \delta n_\nu(\mathbf k)
        =
        -\frac{k_{\mathrm B}T}
        {\varepsilon_\nu^2(\mathbf k)}
        \hbar\mathbf k\cdot\mathbf v.
    \end{equation}
 The correction describes the motion-induced imbalance between opposite incident directions, with branch-dependent weights set by the magnon energies. To relate the occupation imbalance to momentum transfer, we consider scattering within each magnon branch and neglect conversion between the two handednesses. Let $\mathbf k$ and $\mathbf k'$ denote the incoming and outgoing magnon wave vectors, respectively, and let $\chi$ be the angle between them. The differential cross section $d\sigma_\nu(k,\varphi,\chi)/d\chi$ describes scattering into an angular interval $d\chi$ for a magnon in branch $\nu$ with incident wave number $k$ and propagation angle $\varphi$. The momentum transferred to the skyrmion in each event is $\hbar(\mathbf k-\mathbf k')$. Its cross-section-weighted integral is
    \begin{equation}
        \mathbf P_\nu(k,\varphi)
        =
        \int_0^{2\pi}d\chi\,
        \hbar\left[\mathbf k-\mathbf k'_\nu(\chi)\right]
        \frac{d\sigma_\nu(k,\varphi,\chi)}{d\chi}.
    \end{equation}
 We evaluate the scattering for a static skyrmion and assume elastic scattering. The outgoing wave vector therefore satisfies
    \begin{equation}
        \varepsilon_\nu\!\left(k'_\nu,\varphi+\chi\right)
        =
        \varepsilon_\nu(k,\varphi).
    \end{equation}
 For an anisotropic magnon dispersion, energy conservation generally requires an adjustment of the outgoing wave-number magnitude $k'_\nu$. We assume small-angle scattering and neglect the local variation of the magnon dispersion over the small angular interval. Within this approximation, the incoming and outgoing wave vectors have approximately equal magnitudes but different directions,
    \begin{equation}
        \mathbf k'
        \approx
        \left(
        \cos\chi{\mathbf k}
        +
        \sin\chi\,\hat{\mathbf z}\times{\mathbf k}
        \right).
    \end{equation}
 The resulting momentum transfer to the skyrmion is
    \begin{equation}
        \Delta\mathbf p
        \approx
        \hbar(\mathbf k-\mathbf k')
        =
        \hbar 
        \left[
        (1-\cos\chi){\mathbf k}
        -
        \sin\chi\,\hat{\mathbf z}\times{\mathbf k}
        \right].
    \end{equation}
 This approximation neglects only the local change $\varepsilon_\nu(k,\varphi+\chi)-\varepsilon_\nu(k,\varphi)$ associated with a small scattering angle $\chi$. The incident-direction dependence of $\varepsilon_\nu(k,\varphi)$, the thermal occupation, and the group velocity is retained in the integration over $\varphi$, including the branch interchange $\varepsilon_\nu(k,\varphi+\pi/2) =\varepsilon_{-\nu}(k,\varphi)$.

 We define the longitudinal and transverse transport cross sections as
    \begin{equation}
    \begin{aligned}
        \sigma_{\parallel,\nu}(k,\varphi)
        &=
        \int_0^{2\pi}d\chi\,
        (1-\cos\chi)
        \frac{d\sigma_\nu(k,\varphi,\chi)}{d\chi},
        \\
        \sigma_{\perp,\nu}(k,\varphi)
        &=
        -\int_0^{2\pi}d\chi\,
        \sin\chi\,
        \frac{d\sigma_\nu(k,\varphi,\chi)}{d\chi}.
    \end{aligned}
    \end{equation}
 Integrating the momentum transfer over the scattering angle then gives
    \begin{equation}
    \begin{aligned}
        \mathbf P_\nu(k,\varphi)
        &=
        \int_0^{2\pi}d\chi\,
        \Delta\mathbf p\,
        \frac{d\sigma_\nu(k,\varphi,\chi)}{d\chi}
        \\
        &\approx
        \hbar 
        \left[
        \sigma_{\parallel,\nu}(k,\varphi){\mathbf k}
        +
        \sigma_{\perp,\nu}(k,\varphi)
        \hat{\mathbf z}\times{\mathbf k}
        \right].
    \end{aligned}
    \end{equation}
 Using the static scattering cross sections, we weight $\mathbf P_\nu$ by the motion-induced incident flux $\delta n_\nu|\mathbf v_{g,\nu}|$ and integrate over the incident magnon modes,
    \begin{equation}
        \mathbf F_{\mathrm{mag}}
        =
        \sum_{\nu=\pm}
        \int\frac{d^2k}{4\pi^2}\,
        \delta n_\nu(\mathbf k)
        |\mathbf v_{g,\nu}(\mathbf k)|
        \mathbf P_\nu(\mathbf k).
    \end{equation}
 Substituting $\delta n_\nu$ and using $d^2k=k\,dk\,d\varphi$ yields
    \begin{equation}
    \begin{aligned}
        \mathbf F_{\mathrm{mag}}
        ={}&
        -\frac{k_{\mathrm B}T\hbar^2}{4\pi^2}
        \int k\,dk
        \int_0^{2\pi}d\varphi\,
        (\mathbf k\cdot\dot{\mathbf R})
        \\
        &\times
        \left[
        A(k,\varphi)\mathbf k
        +
        B(k,\varphi)\hat{\mathbf z}\times\mathbf k
        \right].
    \end{aligned}
    \end{equation}
 where
    \begin{equation}
    \begin{aligned}
        A(k,\varphi)
        &=
        \sum_{\nu=\pm}
        \frac{
        |\mathbf v_{g,\nu}(k,\varphi)|
        \sigma_{\parallel,\nu}(k,\varphi)
        }{\varepsilon_\nu^2(k,\varphi)},
        \\
        B(k,\varphi)
        &=
        \sum_{\nu=\pm}
        \frac{
        |\mathbf v_{g,\nu}(k,\varphi)|
        \sigma_{\perp,\nu}(k,\varphi)
        }{\varepsilon_\nu^2(k,\varphi)}.
    \end{aligned}
    \end{equation}
 For a scattering response in which a $90^\circ$ rotation interchanges the two chiral branches and reverses the signed transverse deflection, the transport cross sections satisfy
    \begin{equation}
    \begin{aligned}
        \sigma_{\parallel,\nu}(k,\varphi+\pi/2)
        &=
        \sigma_{\parallel,-\nu}(k,\varphi),
        \\
        \sigma_{\perp,\nu}(k,\varphi+\pi/2)
        &=
        -\sigma_{\perp,-\nu}(k,\varphi).
    \end{aligned}
    \end{equation}
 Together with the corresponding interchange of the branch energies and group-speed magnitudes, these relations give, after relabeling $\nu$ as $-\nu$ in the branch sums,
    \begin{equation}
        A(k,\varphi+\pi/2)=A(k,\varphi),
        \qquad
        B(k,\varphi+\pi/2)=-B(k,\varphi).
    \end{equation}
 For the case $\int_0^{2\pi}d\varphi\,B(k,\varphi)\sin2\varphi=0$ in the chosen coordinates, angular integration gives
    \begin{equation}
        \mathbf F_{\mathrm{mag}}
        =
        -2T
        \begin{pmatrix}
            \eta & \lambda_H\\
            \lambda_H & \eta
        \end{pmatrix}
        \dot{\mathbf R},
    \end{equation}
 where
    \begin{equation}
    \begin{aligned}
        \eta
        &=
        \frac{k_{\mathrm B}\hbar^2}{16\pi^2}
        \int k^3\,dk
        \int_0^{2\pi}d\varphi\,A(k,\varphi),
        \\
        \lambda_H
        &=
        \frac{k_{\mathrm B}\hbar^2}{16\pi^2}
        \int k^3\,dk
        \int_0^{2\pi}d\varphi\,
        B(k,\varphi)\cos2\varphi.
    \end{aligned}
    \end{equation}
 Here, the coefficient $\eta$ describes the isotropic magnon drag, while $\lambda_H$ determines the anisotropic coupling between the two velocity components. The effective coefficients $\eta$ and $\lambda_H$ are determined by fitting the theoretical expressions to the simulation data. The scattering cross sections are not independently evaluated here, and the comparison therefore provides a phenomenological description of the observed temperature and angular dependence.

 Defining $\Gamma_{\mathrm{eff}}(T)=s\alpha\mathcal D+\eta T$, the effective Thiele equation \cite{PhysRevLett.30.230} becomes
    \begin{equation}
        \overleftrightarrow{\mathcal M}\ddot{\mathbf R}
        +2\boldsymbol{\mathcal K}
        -2
        \begin{pmatrix}
            \Gamma_{\mathrm{eff}}(T) & \lambda_H T\\
            \lambda_H T & \Gamma_{\mathrm{eff}}(T)
        \end{pmatrix}
        \dot{\mathbf R}
        =0.
    \end{equation}
 The symmetric response is dissipative, with positive definiteness requiring $\Gamma_{\mathrm{eff}}(T)>|\lambda_H T|$. In the steady state and for $|\lambda_H T|\ll\Gamma_{\mathrm{eff}}(T)$, the velocity components are
    \begin{subequations}
        \label{eq24}
    \begin{align}
        v_x(T)
        &=
        \frac{\mathcal K_x}{s\alpha\mathcal D+\eta T}
        -
        \frac{\lambda_H T\mathcal K_y}
        {(s\alpha\mathcal D+\eta T)^2},
        \tag{24a}
        \\
        v_y(T)
        &=
        \frac{\mathcal K_y}{s\alpha\mathcal D+\eta T}
        -
        \frac{\lambda_H T\mathcal K_x}
        {(s\alpha\mathcal D+\eta T)^2}.
        \tag{24b}
    \end{align}
    \end{subequations}
 The off-diagonal terms produce a deflection whose sign depends on the driving direction. For $\boldsymbol{\mathcal K}=\mathcal K(\cos\phi,\sin\phi)$, where $\phi$ is the angle of the driving force relative to the $x$ axis, the signed Hall angle satisfies
    \begin{equation}
        \tan\theta_H
        =
        \frac{v_y\cos\phi-v_x\sin\phi}
        {v_x\cos\phi+v_y\sin\phi}.
    \end{equation}
 To the same order in $\lambda_H T/\Gamma_{\mathrm{eff}}$,
    \begin{equation}\label{eq26}
        \theta_H(\phi,T)
        \approx
        -\frac{\lambda_H T}
        {s\alpha\mathcal D+\eta T}\cos2\phi.
    \end{equation}
 Thus, the Hall angle changes sign between driving along the $x$ and $y$ axes, remains unchanged upon reversal of the driving direction, and vanishes along the diagonal directions.

 Equations~(\ref{eq6}) and (\ref{eq24}) describe the dynamics of an ATM skyrmion at zero and finite temperatures, respectively. To examine these theoretical results, we perform spin simulations based on the stochastic Landau--Lifshitz--Gilbert equation, which is given by~\cite{PhysRevB.107.064423,PhysRevB.99.054423}
    \begin{equation}\label{eq15}
    \frac{\partial{\mathbf{S}_i}}{\partial t} = -\frac{\gamma}{1+\alpha ^2}\mathbf{S}_i \times [(\mathbf{H}_{i}+\mathbf{\zeta}_i)+\alpha \mathbf{S}_i \times(\mathbf{H}_{i}+\mathbf{\zeta}_i)], 
    \end{equation}
 where $\mathbf{H}_{i}= -(1/\mu_s)\partial{{ H}}/\partial{\mathbf{S}_i}$  is the effective field at site $i$ with the magnetic moment $\mu_s$. 
 $\mathbf{\zeta}_i$ denotes a stochastic field modeling the effects of thermal fluctuations.  $\mathbf{\zeta}_i$ has zero mean, and its autocorrelation is given by $\langle\zeta_i^\mu(t)\zeta_j^\nu(t')\rangle = \frac{2\alpha k_{\mathrm B}T}{\mu_s}\delta_{ij}\delta_{\mu\nu}\delta(t-t')$ \cite{PhysRevB.102.104428}, where $\mu$ and $\upsilon$ are Cartesian components and $k_\mathrm B$ is the Boltzmann constant. Unless otherwise specified, we set $K_z=0.4$~meV, $J_1=6$~meV, $J_2=3$~meV, $J_{\mathrm{int}}=-6.6$~meV, $D_0=1$~meV, and $\alpha=0.01$.

\section{\label{sec:level3}RESULTS AND DISCUSSION}

\begin{figure}[t]
        \centering
        \includegraphics[width=0.5\textwidth]{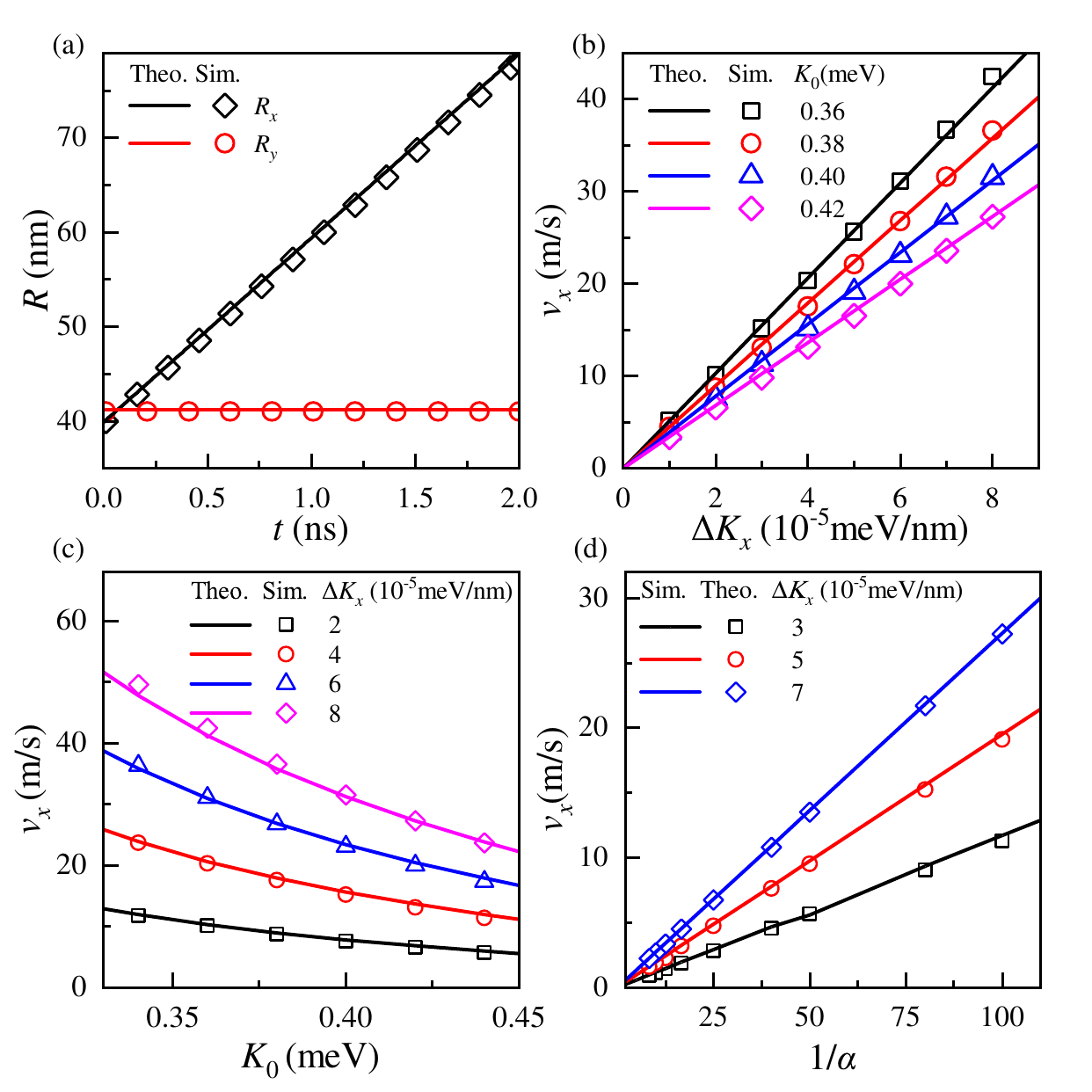}
        \caption{(a) The skyrmion position as a function of time for $K_0 = 0.4$ meV and $\Delta{K}_x= 1.2\times10^{-4}$ meV/nm. The simulated (symbol) and calculated (solid line) skyrmion longitudinal velocities as a function of the (b) anisotropy gradient $\Delta{K}_x$, (c) anisotropy coefficients $K_0$ and (d) the reciprocal of the damping coefficients $1/\alpha$.}
        \label{fig2}
    \end{figure}

 We first establish the zero-temperature transport characteristics of an ATM skyrmion driven by a VCMA gradient. A spatial variation of the magnetic anisotropy makes the skyrmion energy position dependent and therefore produces a force toward the region of lower anisotropy. As shown in Fig.~\ref{fig2}(a), this force drives a steady increase of $R_x$, whereas $R_y$ remains unchanged. The absence of transverse displacement originates from the strong compensation of the gyrotropic response between the two magnetic sublattices. Consequently, the gradient-induced force is balanced primarily by dissipative drag rather than by a transverse force, yielding rectilinear motion. This interpretation is further supported by Fig.~\ref{fig2}(b), where $v_x$ increases linearly with $\Delta K_x$, as expected from the linear dependence of the driving force on the anisotropy gradient. Changing the uniform background anisotropy $K_0$ modifies the skyrmion profile and hence its coupling to the gradient, allowing the mobility to be electrically tuned without introducing a Hall response. The close agreement between the simulations and Eq.~(\ref{eq6}) also shows that the skyrmion remains approximately rigid within the range of gradients considered.

 The dependence on $K_0$ and $\alpha$ reveals how the driving and dissipative sectors separately control the skyrmion velocity. As shown in Fig.~\ref{fig2}(c), increasing $K_0$ reduces $v_x$ for every $\Delta K_x$. A stronger easy-axis anisotropy reshapes and generally localizes the skyrmion texture, thereby reducing the texture integral $\mathcal K$ that converts the anisotropy gradient into a translational force. A larger $\Delta K_x$ nevertheless enhances the force and shifts the velocity curves upward. By contrast, Gilbert damping does not change the origin of the driving force but controls the rate at which the driving energy is dissipated. The linear dependence $v_x\propto1/\alpha$ in Fig.~\ref{fig2}(d) therefore directly reflects the steady-state force balance $v_x=\mathcal K_x/(s\alpha\mathcal D)$, where $\mathcal D$ characterizes the dissipative weight of the skyrmion texture. The simultaneous agreement with theory in Figs.~\ref{fig2}(b)--\ref{fig2}(d) demonstrates that the observed parameter dependence is governed by the equilibrium texture and its collective coefficients, rather than by gradient-induced deformation. 

 Having established the Hall-free ATM skyrmion dynamics at zero temperature, we now turn to the finite temperature regime, where thermally excited magnons provide additional channels for momentum exchange between the magnetic background and the skyrmion. The resulting backaction is governed by the handedness-resolved magnon spectra, which determine the thermal occupation and transport weight of each branch. To characterize the magnon excitations, we linearize the spin dynamics around the collinear ground state $\mathbf{S}^{A}=\hat{\mathbf{z}}$ and $\mathbf{S}^{B}=-\hat{\mathbf{z}}$. Within linear spin-wave theory, the two sublattices are described by
    \begin{align}
    A_{\mathbf{k}}^{A}
    &=
    -J_{\rm int}+2K_z
    +2J_1[1-\cos(k_xa)]
    +2J_2[1-\cos(k_ya)], \\
    A_{\mathbf{k}}^{B}
    &=
    -J_{\rm int}+2K_z
    +2J_2[1-\cos(k_xa)]
    +2J_1[1-\cos(k_ya)] .
    \end{align}
 The two magnon branches are then obtained as
    \begin{equation}
    s\omega_{\pm}(\mathbf{k})
    =
    \sqrt{\bar A_{\mathbf{k}}^2-J_{\rm int}^2}
    \pm\delta_{\mathbf{k}},
    \label{eq:magnon_dispersion}
    \end{equation}
 where
    \begin{align}
    \bar A_{\mathbf{k}}
    &=
    -J_{\rm int}+2K_z
    +(J_1+J_2)
    \left[
    2-\cos(k_xa)-\cos(k_ya)
    \right], \\
    \delta_{\mathbf{k}}
    &=
    (J_1-J_2)
    \left[
    \cos(k_xa)-\cos(k_ya)
    \right].
    \end{align}
 Here, $\delta_{\mathbf{k}}$ gives the characteristic $d$-wave altermagnetic magnon splitting. Consequently, the two modes are degenerate along $k_x=\pm k_y$, including the $\Gamma$--$\rm M$ direction, whereas a finite splitting appears along the $\Gamma$--$\rm X$ and $\Gamma$--$\rm Y$ directions. The splitting is
    \begin{equation}
    \Delta\omega_{\mathbf{k}}
    =
    \frac{2}{\hbar}
    |J_1-J_2|
    \left|
    \cos(k_xa)-\cos(k_ya)
    \right|.
    \end{equation}
 
    \begin{figure}[t]
        \centering
        \includegraphics[width=0.5\textwidth]{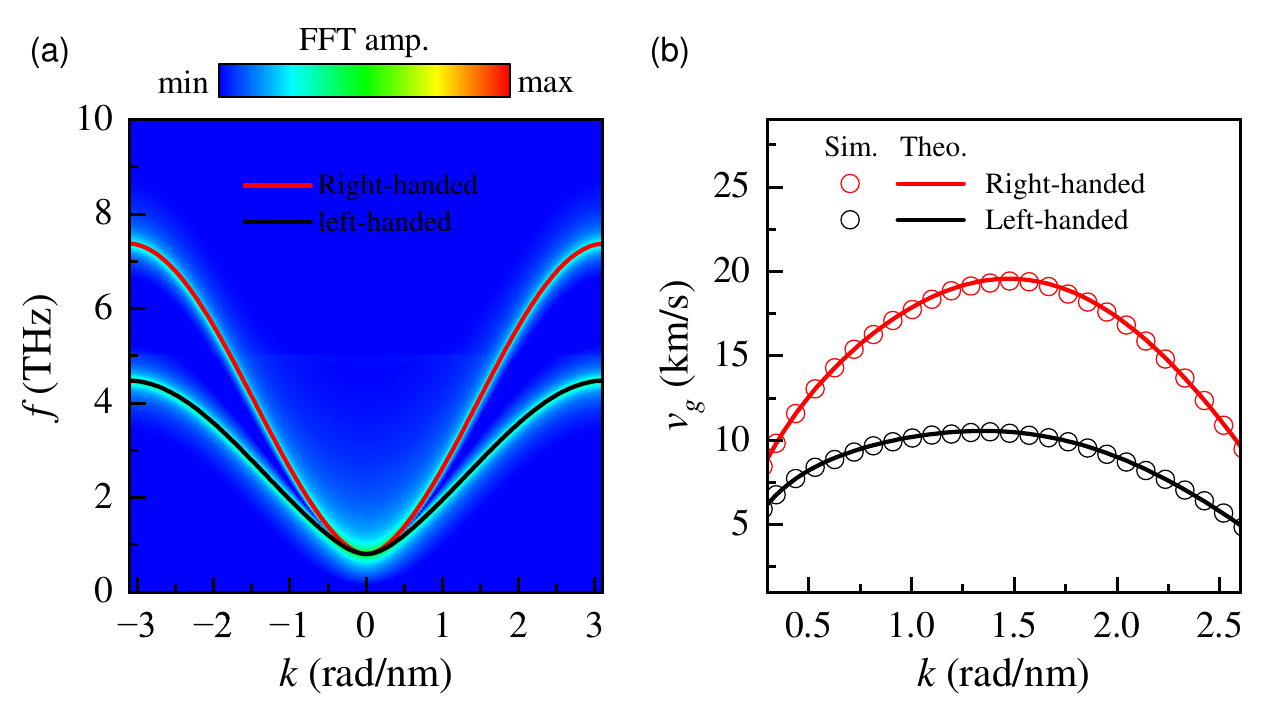}
        \caption{(a) Dispersion and (b) group velocity of altermagnetic magnons propagating along the $x$ direction. In (a), the color map shows the Fourier amplitude of the simulated spin dynamics, while the solid curves represent the theoretical dispersions obtained from Eq.~(\ref{eq:magnon_dispersion}). In (b), the open circles and solid curves denote the group velocities obtained by numerically differentiating the simulated dispersions and analytically differentiating Eq.~(\ref{eq:magnon_dispersion}), respectively. Red and black correspond to right- and left-handed magnons, respectively. The parameters are $J_1=6$ meV, $J_2=3$ meV, $D_0=1$ meV, $J_{\mathrm{int}}=-6.6$ meV, and $K_0=0.4$ meV.}
        \label{fig3}
    \end{figure}
 
 To obtain these spectra, we perform a spatiotemporal Fourier transform of the simulated spin configurations and extract the magnon branches from the spectral peaks in the $(k,\omega)$ space. Figure~\ref{fig3}(a) shows the dispersions of the two magnon branches with opposite handedness propagating along the $x$ direction, together with the theoretical results from Eq.~(\ref{eq:magnon_dispersion}). The corresponding group velocities, $v_{g,\nu}(k)=\partial\omega_\nu(k)/\partial k$, are obtained by numerically differentiating the simulated dispersions and analytically differentiating Eq.~(\ref{eq:magnon_dispersion}), respectively. As shown in Fig.~\ref{fig3}(b), the two branches become increasingly distinct away from the band minimum, exhibiting different dispersions and group velocities. In particular, the right-handed branch exhibits a larger group velocity than the left-handed branch over most of the wave-vector range considered. The handedness-dependent dispersions therefore generate unequal branch-resolved phase-space and transport weights.

 Consequently, the oppositely signed transverse momentum transfers associated with the two magnon handednesses do not cancel exactly. Although the two branches are skew scattered in opposite transverse directions, their contributions generally differ in magnitude because their thermal occupations, group velocities, and scattering cross sections are not identical. This incomplete cancellation produces a finite net transverse momentum transfer from the magnon bath to the skyrmion, providing the microscopic origin of the thermally induced transverse force. It thereby opens a Hall-response channel that is absent at zero temperature. Within the classical approximation, the magnon occupations increase with temperature, strengthening the magnon backaction on the moving skyrmion.
 
    \begin{figure}[t]
        \centering
        \includegraphics[width=0.5\textwidth]{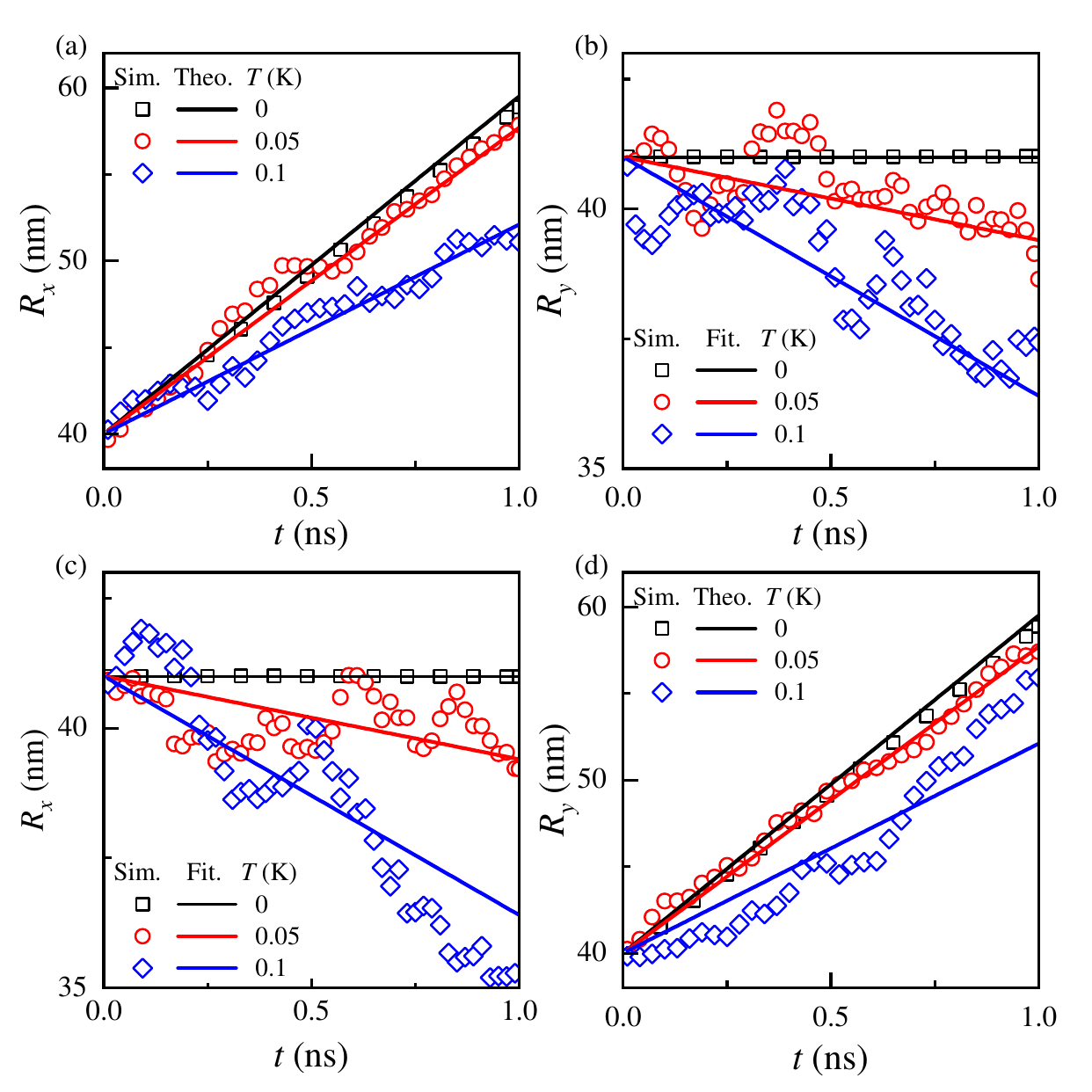}
        \caption{Skyrmion (a) longitudinal position $R_x$ and (b) transverse position $R_y$ as functions of time for anisotropy gradient along the $x$ direction at different temperatures. Skyrmion (c) transverse position $R_x$ and (d) longitudinal position $R_y$ as functions of time for anisotropy gradient along the $y$ direction. The trajectories are averaged over five independent runs with different random seeds. In simulations, we set $K_0=0.4$ meV and $\Delta K_\xi=1.2 \times10^{-4}$ meV/nm.}
        \label{fig4}
    \end{figure}

    \begin{figure}[t]
        \centering
        \includegraphics[width=0.5\textwidth]{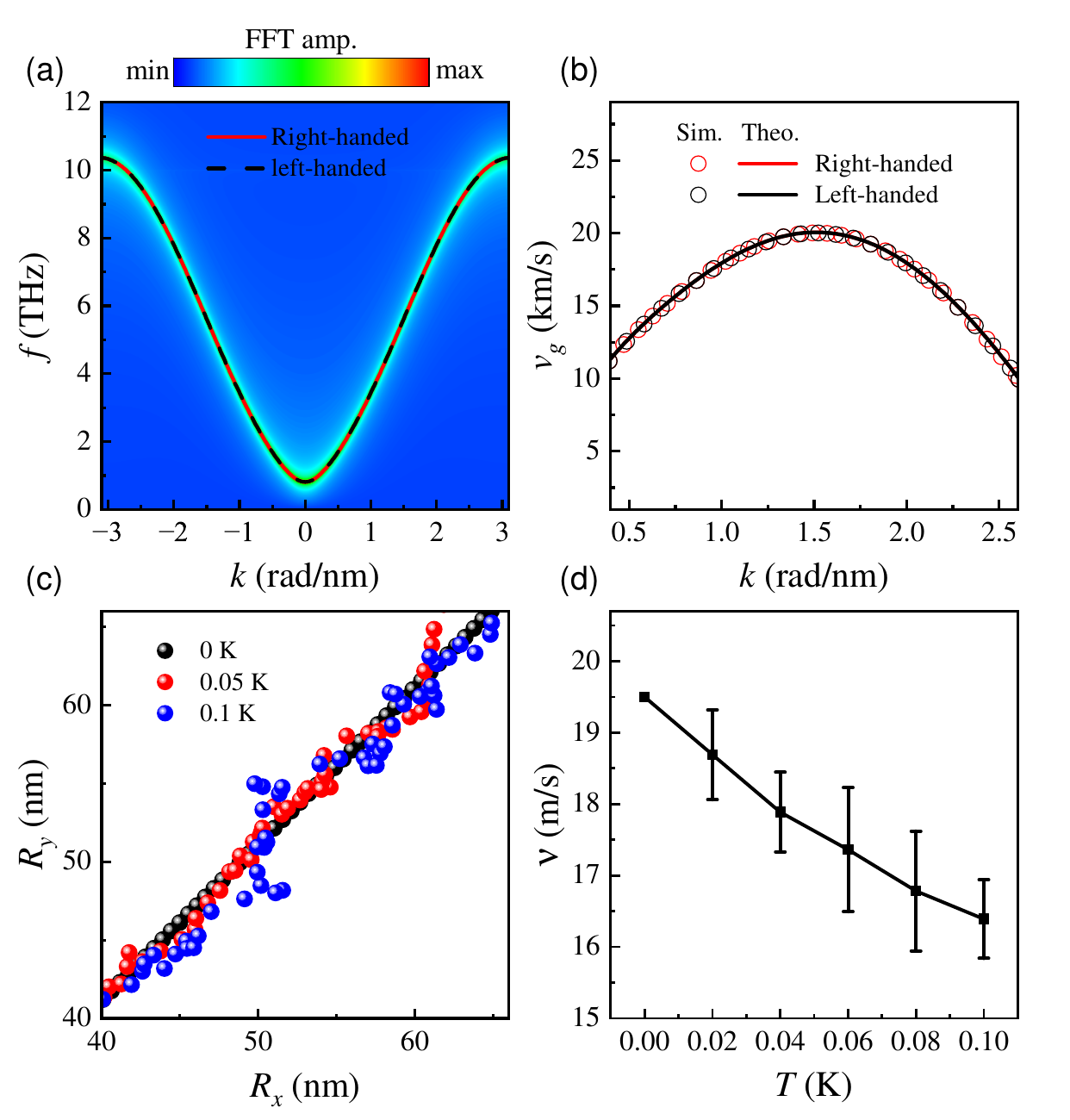}
        \caption{(a) Magnon dispersion and (b) group velocity along the high-symmetry $\Gamma\rightarrow\mathrm{M}$ direction. In (a), the color map shows the Fourier amplitude of the simulated spin dynamics, and the red solid and black dashed curves are the theoretical results for the two magnon handednesses. In (b), open circles and solid curves denote the simulation and theoretical results, respectively. (c) Skyrmion trajectories in the $R_x$--$R_y$ plane for an anisotropy gradient along $\Gamma\rightarrow \text M$ at $T$ = 0, 0.05, and 0.1 K. The trajectories are averaged over five independent runs with different random seeds. (d) Skyrmion velocity as a function of temperature. The symbols represent the mean velocities, and the error bars span the minimum and maximum values across the five runs. The parameters are $K_0=0.4$~meV and $\Delta K=1.2\times10^{-4}$~meV/nm.}
        \label{fig5}
    \end{figure}

 The temperature-dependent skyrmion motion is shown in Fig.~\ref{fig4}. For an anisotropy gradient applied along the $x$ direction [Figs.~\ref{fig4}(a) and \ref{fig4}(b)], the longitudinal motion slows down as the temperature increases. At finite temperature, thermally excited magnons exchange momentum with the moving skyrmion and provide an additional dissipative channel, giving rise to the magnon drag $\eta T$ in $\Gamma_{\mathrm{eff}}(T)$. Consequently, the slope of $R_x(t)$ decreases progressively with increasing temperature. Meanwhile, a transverse drift develops along the negative $y$ direction. This transverse response originates from the directional imbalance of the two magnon handednesses. Because their dispersions and group velocities are split along the principal axes, their branch-resolved thermal fluxes and skew-scattering momentum transfers do not cancel completely. The remaining anisotropic contribution is described by $\lambda_H$, which couples the longitudinal motion to a transverse dissipative force and yields $v_y<0$ for driving along $+\hat{x}$.

 For a gradient applied along the $y$ direction [Figs.~\ref{fig4}(c) and \ref{fig4}(d)], the same two effects are observed. The longitudinal velocity along $y$ decreases with temperature because of the enhanced thermal-magnon drag, whereas the uncompensated handedness-dependent momentum transfer produces a transverse motion along the negative $x$ direction. The corresponding $v_x<0$ is consistent with Eq.~(\ref{eq24}). Although the Cartesian transverse components are negative for both principal driving directions, the signed deflection relative to the driving force reverses between the $x$ and $y$ directions, reflecting the $d$-wave anisotropy of the magnon response.

 A particularly instructive case occurs when the anisotropy gradient is oriented along the $\Gamma\rightarrow \mathrm{M}$ direction of the Brillouin zone, with $\mathrm{M}=(\pi/a,\pi/a)$. Along this high-symmetry path, the dispersions of the two magnon handednesses remain degenerate~\cite{PhysRevB.108.L180401}, as shown in Fig.~\ref{fig5}(a). Their group velocity components along this path are therefore identical, consistent with Fig.~\ref{fig5}(b). Within the effective Thiele equation Eq. (23), the diagonal directions are eigenvectors of the response matrix. The skyrmion velocity therefore remains parallel to the driving force along these directions, yielding a vanishing Hall angle. This Hall-free motion is directly reflected in the skyrmion trajectories shown in Fig.~\ref{fig5}(c). When the anisotropy gradient is applied along the $\Gamma\rightarrow \mathrm{M}$ direction, the skyrmion follows the driving direction at $T=0$, $0.05$, and $0.1$~K without developing a systematic transverse deflection. Increasing temperature enhances the stochastic fluctuations but does not generate a finite Hall angle. The longitudinal motion, however, becomes slower with increasing temperature, as shown in Fig.~\ref{fig5}(d), where the skyrmion velocity decreases from approximately $19.5$ to $16.3$~m/s between $T=0$ and $0.1$~K. This reduction originates from the enhanced magnon-induced friction at finite temperature. Therefore, along the $\Gamma\rightarrow \mathrm{M}$ direction, the transverse forces from the two magnon handednesses remain compensated, while their longitudinal dissipative contributions lead to a temperature-dependent reduction of the skyrmion velocity.

     \begin{figure}[t]
        \centering
        \includegraphics[width=0.5\textwidth]{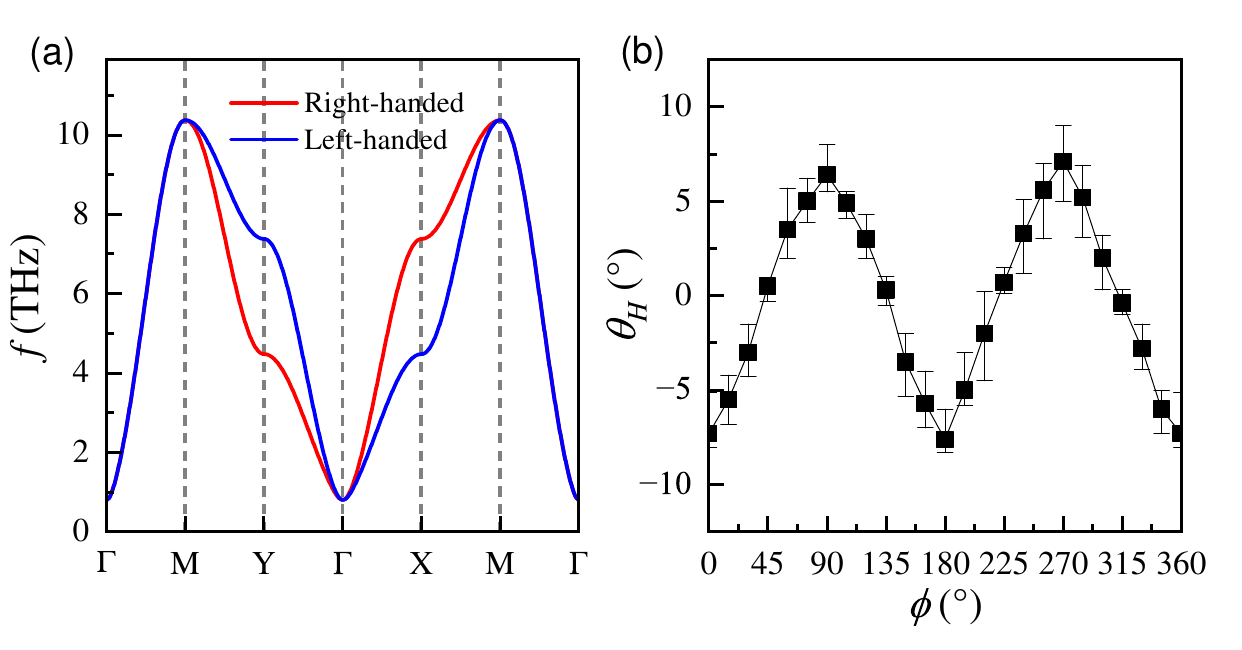}
        \caption{(a) Magnon band structure of the altermagnet plotted along a Brillouin zone path in a $k_x-k_y$ plane crossing the $\Gamma$ point. The red and blue curves represent the right- and left-handed magnon branches, respectively. (b) Angular dependence of the skyrmion Hall angle $\theta_H$ on the driving direction $\phi$ at $T=0.02~\text K$. The symbols represent the mean Hall angles over five independent runs with different random seeds, and the error bars span the minimum and maximum values. }
        \label{fig6}
    \end{figure}

 To further verify the angular dependence predicted by Eq.~(\ref{eq26}), we examine the magnon dispersions and systematically vary the direction of the anisotropy gradient, as shown in Fig.~\ref{fig6}. Figure~\ref{fig6}(a) shows that the two magnon handednesses exhibit distinct dispersions along the $\Gamma$--X and $\Gamma$--Y directions, with their branches interchanged under a $90^\circ$ rotation. In contrast, the two branches remain degenerate along the diagonal $\Gamma$--M direction. The corresponding skyrmion Hall angle exhibits an approximately $\cos 2\phi$ dependence [Fig.~\ref{fig6}(b)], consistent with Eq.~(\ref{eq26}). Finite Hall deflections occur along directions with magnon chiral splitting, with opposite signs for driving along the $x$ and $y$ axes. Along the diagonal directions, the Hall angle vanishes due to the cancellation of the transverse contributions from the two magnon handednesses. Moreover, the Hall angle reverses sign under a $90^\circ$ rotation but remains unchanged under a $180^\circ$ rotation of the driving direction. This angular dependence is consistent with the $\mathcal C_4\mathcal T$ symmetry of the altermagnetic magnon spectrum, further supporting the magnon-induced origin of the skyrmion Hall effect. Experimentally, quantum relaxometry using nitrogen-vacancy (NV) centers in diamond has been proposed as a noninvasive probe of anisotropic spin transport in altermagnetic insulators~\cite{2ppn-kvjv}. Combined with the demonstrated capability of NV magnetometry to image skyrmion spin textures~\cite{Dovzhenko2018}, these advances suggest a potential route toward probing the connection between magnon transport and skyrmion dynamics in altermagnets.

\section{\label{sec:level4}CONCLUSION}

 In summary, we have demonstrated that a VCMA gradient provides a current-free route for driving ATM skyrmions. In the absence of thermal fluctuations, altermagnetic symmetry enforces a vanishing transverse response, resulting in rectilinear skyrmion motion. The longitudinal velocity increases with the anisotropy gradient, scales inversely with the Gilbert damping, and can be further controlled by the background magnetic anisotropy through its influence on the equilibrium skyrmion profile and the associated collective coefficients. Upon thermal excitation, magnon--skyrmion scattering introduces additional longitudinal drag and generates a transverse reaction force through handedness-dependent skew scattering. Owing to the anisotropic altermagnetic magnon band structure, the relative transport weights of the two magnon handednesses are interchanged between propagation along the $x$ and $y$ directions, resulting in skyrmion Hall angles of opposite sign for the two driving directions. By contrast, along the high-symmetry direction, the two magnon modes remain degenerate, so that their oppositely signed transverse momentum transfers cancel and Hall-free motion persists even at finite temperature. These findings establish altermagnets as a promising platform for thermally tunable skyrmion transport. Since the underlying mechanism originates from the intrinsic anisotropy of the altermagnetic magnon spectrum, the direction-dependent Hall response is expected to persist under other driving mechanisms and may serve as an indicator of altermagnetic order.

\begin{acknowledgments}
 The work is supported by the Natural Science Foundation of China (Grants No. U22A20117, and No. 52371243, and No. 12404270), Guangdong Provincial Quantum Science Strategic initiative (Grant No. GDZX2401002), the Natural Science Foundation of Shandong Province (Grants No. ZR2025QC13, and NO. ZR2024QA166), the Guangdong Basic and Applied Basic Research Foundation (Grant No. 2024A1515012665), and the Liaocheng University Start-up Fund for Doctoral Scientific Research (Grant No. 318052379).
\end{acknowledgments}

\section*{DATA AVAILABILITY}
 The data that support the findings of this article are available from the authors upon reasonable request.


\end{document}